\documentclass[aps,twocolumn,prl,amsmath,amssymb,floatfix]{revtex4-2}
\usepackage{graphicx}
\makeatletter

\begin{document}
\title{A theorem of Kohn applied to quantum oscillations in Cuprates}

\author{Sudip Chakravarty}
\affiliation{Mani L. Bhaumik Institute for Theoretical Physics\\Department of Physics and Astronomy, University of
California Los Angeles, Los Angeles, California 90095-1547}

%\date{\today}
\begin{abstract}
In this note I apply a theorem of Kohn to quantum oscillations in cuprates. I show that when combined with Gell-Mann and Low theorem,
there is rigorous justification of quantum oscillations as in a Fermi liquid  for cuprates. The Gell-Man-Low theorem is not perturbative, and applies more generally as long as the adiabatic evolution from an isolated non-degenerate state   is the case. Oscillation frequencies are identical to the non-interacting problem.  Thus, quantum oscillations protect the Fermi liquid and is in turn protected by it. This is not to imply that  other properties of the cuprates that do not have a gap in the spectrum could not exhibit non-Fermi liquid behavior, for example, angle resolved photoemission spectroscopy.
\end{abstract}
\maketitle
In a beautiful set of experiments  quantum oscillations in a very high magnetic field were discovered in cuprates; for a recent review, see Sebastian and Proust~\cite{Sebastian:2015}. While much discussions took place during the ensuing years since their discoveries, the initial optimism~\cite{Chakravarty:2008} that the field of high temperature superconductivity was simplified with the discovery of quantum oscillations was largely dashed; the history is not worth reviewing here. I return to this optimistic view once more  and show how optimism can be restored.

The problem was numerous appeals to non-Fermi liquid (nFL) phenomenology that indeed appeared tantalizing. There is no proof, or a good argument as to why  nFL should have Landau levels and hence quantum oscillations. By a non-Fermi liquid, I mean the  lack of quasiparticle poles of the Green function on the second Riemann sheet. Without quasiparticles, how could vector potental couple correctly to the quasiparticle and give rise to Landau levels? There is something amiss! This perplexed the present author sometime ago.~\cite{Chakravarty:2011} Over the years I have come to realize that the resolution of this apparent paradox is not that difficult to understand: (i) a number of properties, which do not involve gaps in the  excitation can be driven to a strongly interacting nFL state, and may be that is the case  for angle resolved photoemission spectroscopy;~\cite{Damascelli:2003,Abrahams:2000} (ii) on the other hand, for quantum oscillations in a fully filled Landau level involves  a isolated non-degenerate state separated from the excited state by a gap. It is this gap that protects the quantum oscillations as in a Fermi liquid and in turn is protected by it. There is no conflict between  the two ideas here.

One often wonders about the half-filled lowest Landau level being a rather complex system composed of composite fermions.~\cite{Jain:2007} And indeed there is some unusual evidence of Schubnikov-de Haas-like  oscillations as one moves away from the half-filled limit.~\cite{Du:1994} But clearly the half-filled limit is highly degenerate and there is no protection  against  adiabaticity.~\cite{Halperin:1993}

To  appreciate the problem, it is extremely important to revisit  a theorem  of Kohn~\cite{Kohn:1961} that states that in a two dimensional continuum system electron-electron interaction cannot shift the dHvA (de Haas-van Alphen)  frequency. So, consider a continuum two-dimensional electronic system with arbitrary short ranged  velocity independent electron-electron interaction, but without disorder. Velocity dependent interaction can considerably change the picture, however.~\cite{Holstein:1973,Chakravarty:1995} We return to the question of disorder later. The Hamiltonian is

\begin{equation}
\begin{split}
{\cal H}&= \int_{\mathbf x} \psi_{\sigma}^{\dagger}({\mathbf x})\varepsilon({\mathbf p}, {\mathbf x}) \psi_{\sigma}({\mathbf x}) + \\ &\int_{{\mathbf x},{\mathbf x'}}\psi_{\sigma}^{\dagger}({\mathbf x})\psi_{\sigma'}^{\dagger}({\mathbf x'}) v({\mathbf x}- {\mathbf x'})\psi_{\sigma'}({\mathbf x'})\psi_{\sigma}({\mathbf x})
\end{split}
\end{equation}
Here $\varepsilon({\mathbf x},{\mathbf p})$ is the Hamiltonian for individual particles of mass $m$, charge $e$, magnetic moment $\mu_{B}$, spin $\frac{1}{2}$, moving in an external magnetic field $\mathbf B$. Here $v({\mathbf x}- {\mathbf x'})$ is assumed to be a spin independent static potential, so any explicit spin dependence can be dropped. Moreover, for simplicity we shall also drop the Zeeman term $-g\mu_{B}{\mathbf B}\cdot \boldsymbol{\sigma}$ in the Hamiltonian. The notation $\int_{\bf x}$ stands for spatial integration. 

The single particle energy is 
\begin{equation}
\varepsilon({\mathbf p},{\mathbf x}) = \frac{1}{2m}\left[ {\mathbf p} - (e/c) {\mathbf A}\right]^{2}.
\end{equation}
 Note that since magnetic field is {\em never} a small perturbation, as it increases indefinitely  with the size of the system, it is  imperative  to begin with the noninteracting system but in the presence of the magnetic field and then consider the effect of the interaction term. 

Let the unperturbed problem be defined by the non-interacting Hamiltonian, $H_{0}$, in a rectangular box $L_{x}\times L_{y}$. For simplicity we shall consider  two-dimensions and come back later to add the third dimension. In the Landau gauge for the magnetic field $B\hat{z}$ ($c$ being the velocity of light),
\begin{equation}
H_{0}=\frac{1}{2m}\sum_{i}\left[p_{x,i}^{2}+\left(p_{y,i}+\frac{e}{c}Bx_{i}\right)^{2}\right]
\end{equation} 
The solution of this Landau level problem is of course   textbook material. The energy eigenvalues and eigenfunctions are 
\begin{equation}
\epsilon_{n,{\bf k}}=\hbar\omega_{c}\left(n+\frac{1}{2}\right), \; \psi_{n, k}= e^{iky}u_{n}\left(x+\frac{\hbar c k}{eB}\right),
\end{equation}
($u_n$ is the harmonic oscillator wave function) where the frequency $\omega_{c}=\frac{eB}{mc}$ and the degeneracy of each energy level is
\begin{equation}
d_{\Phi}=2\frac{\Phi}{\Phi_{0}} ,
\end{equation}
where the total flux threading the system is $\Phi= B L_{x}L_{y}$ and the flux quantum is $\Phi_{0}=hc/e$; the factor of 2 is for spin. The functions $u_{n}$ are the one-dimensional harmonic oscillator wave functions.

Even though momenta are not good quantum numbers, it is still useful to visualize the spectra on the two-dimensional $k_{x}-k_{y}$-plane. The degenerate spectra, of degeneracy $g$, lie in concentric circles in this plane, separated by $\hbar\omega_{c}$. Since momentum is no longer a good quantum number, the states are not located on specific points on the circle but can be viewed as rotating with frequency $\omega_{c}$.  In particular, if we denote $\Delta A$ to be the area between the concentric circles, then
\begin{equation}
\frac{L_{x}L_{y}}{(2\pi)^{2}}\Delta A=\frac{d_{\Phi}}{2}.
\label{eq:degeneracy}
\end{equation}
Imagine that the Fermi level, $\epsilon_{F}$, at $T=0$ is situated on one such concentric level such that all states with energy $E\le  \epsilon_{F}$ are completely  filled and all the levels for $E> \epsilon_{F}$ are completely empty. Then the total  number of occupied states is {\em eaxctly} the same as the system without the magnetic field and the total energy per electron is also {\em exactly} the same. The area enclosed by the Fermi level, $A(\epsilon_{F})$, follows trivially from Eq.~(\ref{eq:degeneracy}) and is
\begin{equation}
2\frac{A(\epsilon_{F})}{(2\pi)^{2}} = \frac{N}{L_{x}L_{y}},
\end{equation}
where $N$ is the total number of particles. That is none other than the Luttinger sum rule~\cite{Luttinger:1960}. It is important to note that even though a magnetic field is never a small perturbation, the Luttinger sum rule is unchanged.

The magnetic field corresponding to the ground state of a  system with  an integer number, $n(\epsilon_{F})$, of Landau levels completely filled and all the rest completely empty will satisfy
\begin{equation}
\frac{1}{B_{n}}= n(\epsilon_{F})\frac{2\pi e}{\hbar c}\frac{1}{A(\epsilon_{F})}.
\label{eq:Bn}
\end{equation}
Note that  this ground state is an isolated nondegenerate state separated by a {\bf gap}  from the excited state. As we increase $B$, the quantized orbits are drawn out of the Fermi level, and sequentially pass through {\em essentially} identical set of nondgenerate isolated ground states. This is certainly true when $n\gg 1$. For low Landau levels the assumption is not  correct. However, the case of interest is higher Landau levels. This of course results in the periodicity in the properties of the electron gas; the correction in the limit that $n(\epsilon_{F})\gg 1$ is negligible. Periodicity of course does not imply sinusoidal wave form and can contain higher harmonics.

Now, fix $B_{n}$ to a completely filled Landau level and turn on the electron-electron interaction. By {\em adiabatic continuity} a nondegenerate isolated ground state will remain nondegenerate and therefore the sequence of states corresponding to fully filled Landau levels as a function of the magnetic field will be the same, as in the noninteracting case. The periodicity is therefore unchanged and is determined by the enclosed area $A(\epsilon_{F})$, which in turn is fixed by the Luttinger sum rule. 

The third dimension can be incorporated by considering the energy spectrum as a function of the $z$-component of the wave number $k_z$,  $\epsilon_n(k_z)$, with no added complications. We just have to repeat the argument for a fixed $k_z$

The argument will clearly break down if electron-electron interaction collapses the gap, but in that case there is no quantum oscillations at all.   Note that Kohn's theorem makes no statement about the amplitude nor about  about the waveform of the oscillations. In general the periodicity, when Fourier analyzed will contain harmonics.

The adiabatic argument can be quantified with the help of Gell-Mann and Low theorem. let $|\Psi_0\rangle$ be an eigenstate, not necessarily the ground state, of $H_0$ with energy $E_0$ and let the interacting Hamiltonian $H=H_0+gV$ where $g$ is the coupling constant and $V$ is the interaction term. We define a Hamiltonian $H_t=H_0+e^{-\epsilon|t|}gV$ which interpolates between $H$ and $H_0$ in the limit $\epsilon \to 0^+$ and $|t|\to\infty$. 

Let $U_{\epsilon I}$ denote the evolution operator in the interaction picture. The Gell-Mann and Low theorem~\cite{Gell-Mann:1951} asserts that if the limit $\epsilon\to 0^+$ of 
\begin{equation}
|\Psi_\epsilon^\pm\rangle = \frac{U_{\epsilon I}(0\pm\infty)|\Psi_0\rangle}{\langle \Psi_0| U_{\epsilon  I}(0\pm\infty)|\Psi_0\rangle}
\end{equation}
exists then $|\Psi_\epsilon^\pm\rangle$ are eigenstates of $H$. 
The original proof is  perturbative, so are many subsequent proofs.~\cite{Fetter:1971} The non-perturbativr proof by Molinari~\cite{Molinari:2007} is recent; see also Ref.~\cite{Nenciu:1989}. Applied to the ground state one cannot rule out level crossings. This would mean quantum quantum phase transition. Avoided level crossing as the volume tends to infinity will signify quantum critical points. There is no evidence for either in experiments. Perhaps none of these complications arise because the original ground state in the present context is isolated and non-degenerate. This theorem completely confirms the original statement of Kohn in the context of quantum oscillations.

An important point here is the gap between the filled Landau levels and the empty ones, which protects the Fermi liquid character.  So, the phenomenology can be understood in the conventional language.  There is no need to invoke non-Fermi liquid phenomenology unless the gap collapses, in which case there would be no quantum oscillations, as stated earlier. {\em This is not to imply that interaction driven non-Fermi liquids---non-existence of quasiparticles---cannot  exist for properties other than quantum oscillations, where there are no gaps at the Fermi surface. It is only that quantum oscillations protect the Fermi liquid and is in turn protected by it.} Scaleless (critical) Fermi surface is a strange object where the modes conventionally decouple unless there is a mechanism that opens up gaps---a  phenomenon that reminds us of the Kondo problem.

In an often quoted paper by Luttinger~\cite{Luttinger:1961} a derivation of the Lifshitz-Kosevich (LK)  formula for dHvA is given. However,  LK formula is valid {\em if and only if} the Fermi liquid theory is valid, which is what Kohn's theorem protects. As Kohn correctly points out,~\cite{Kohn:1961} it is not the analyticity, but continuity that is responsible for quantum oscillations in the interacting  system. If the LK formula holds, we are very likely observing quasiparticles, perhaps with renormalized masses, in complete agreement with the experimentalists~\cite{Sebastian:2015}.

It is useful to recapitulate what Luttinger~\cite{Luttinger:1961} really proved. The parameters of interest are $k_{B}T, \; \hbar\omega_{c} \ll \mu$ and $k_{B}T$  not much larger than $\hbar\omega_{c}$.  He separated the self energy $\Sigma$   into three parts:
\begin{equation}
\Sigma=\Sigma_{0}+\Sigma_{T}+\Sigma_{osc}.
\end{equation}
 Here $\Sigma_{0}$ is a {\em field independent} part  at $T=0$ and $\Sigma_{T}$ is the first temperature correction, which by the Sommerfeld expansion is 
 \begin{equation}
 \Sigma_{T}\sim (k_{B}T/\mu)^{2}\Sigma_{0}.
 \end{equation}
Moreover  Luttinger estimates that
\begin{equation}
\Sigma_{osc}\sim (\hbar\omega_{c}/\mu)^{3/2},
\end{equation}
and therefore 
 \begin{equation}
 \Sigma_{osc}/\Sigma_{T}\sim (\hbar\omega_{c}/k_{B}T)^{2} (\mu/\hbar\omega_{c})^{1/2} \gg 1,
 \end{equation}
 allowing us to drop $\Sigma_{T}$.
 The  assumption here is that ts  Sommerfeld expansion is a meaningful asymptotic expansion. It may ``fail" in two-dimensions. These estimates allow him to drop $\Sigma_{T}$. If we now assume that the electron-electron scattering rate  vanishes as $(\epsilon-\mu)^{2}$, which is a {\em Fermi liquid assumption}, the leading oscillatory part of the thermodynamic potential is
 \begin{equation}
 \Omega_{osc}= - \frac{1}{\beta} \sum_{r}\ln \left[ 1 +e^{\beta(\mu - E_{r})}\right],
 \end{equation}
 which is exactly the thermodynamic potential for independent fermions but with the renormalized quasiparticle energies $E_{r}$ determined from 
 \begin{equation}
E_{r} - Q_{r}(E_{r})=0,
\end{equation}
where $Q_{r}$ is the real part of the self energy.
 Except for that it is exactly the LK formula.
 
 The final formula of Luttinger can of course be cast in terms of fermionic  Matsubara frequencies, $\omega_{n}$,  and a self energy dependent on it. It then reads
 \begin{equation}
 \Omega_{osc}=-\frac{1}{\beta}\sum_{n} \mathrm{Tr}\{\ln\left[\epsilon(\bf B)+\Sigma_{0}(\omega_{n})-i\omega_{n}\right]\}.
 \end{equation}
Here $\epsilon(\bf B)$ is unperturbed single particle Hamiltonian.  It appears that the non-oscillatory part of the self energy $\Sigma_{0}(\omega_{n})$  can be phenomenologically assigned  whatever we wish, in particular a non-Fermi liquid or a marginal Fermi liquid form.~\cite{Abrahams:2000} This procedure~\cite{Wasserman:1991,Pelzer:1991,McCollam:2008} seriously lacks consistency, as there is no proof that nFLs exhibit Landau levels, which  is the {\em sine qua non} of quantum oscillations.
 
 Impurities will break translational invariance and Kohn's theorem cannot, strictly speaking, hold. A moderate amount of disorder can be understood in terms of a self consistent Born approximation and has been discussed extensively~\cite{Ando:1974}. Landau levels will be broadened and will overlap, but as long as the states within a Landau band are not fully localized, magnetic oscillations should persist. Can we understand, on dimensional grounds, how the oscillations are affected by disorder. Impurity broadening must lead to the decay of the amplitudes characterized by the Dingle factors. How about the frequency? With disorder we have a new dimensionless parameter, $\hbar/(\epsilon_{F}\tau)$, that is expected to affect the frequency as well. The electron will take longer to complete a cyclotron orbit, so the frequency should be shifted downward, but by what amount? The downward shift in the energy $\epsilon$ of an extremal orbit can be estimated to be
\begin{equation}
\Delta(\epsilon) = \frac{P}{\pi}\int d\omega \frac{\Gamma(\epsilon,\omega)}{\epsilon-\omega},
\end{equation}
where $\Gamma(\epsilon_{k},\omega)=\pi \sum_{k\ne k'}|V_{k,k'}|^{2}\delta(\epsilon_{k'}-\omega)$ and $\epsilon_{k}=\epsilon=\epsilon_{F}$. When averaged over the distribution of disorder, $\Gamma$ is smooth and independent of energy, and therefore the principal value integral vanishes. A more refined self consistent argument~\cite{Goswami:2008} shows that the correction to the frequency is of order $(\hbar/\epsilon_{F}\tau)^{2}$, which is a small correction in most cases.

It is important to summarize the line of reasoning in this paper. Kohn's theorem leads us to argue  that the quantum oscillations in an interacting system, in particular dHvA oscillations, are adibatically continuous to the non intearcting problem with identical frequencies! The crux is the gap, $\hbar \omega_c$, between the fully filled Landau levels and the empty ones. This  gap, in addition to a unique non-degenerate unperturbed ground state, justify the adiabatic continuity.via Gell-Mann and Low theorem, completely substantiating the original argument of Kohn, We also learn thet nFL ideas are not necessary in understanding quantum uscillations.  This is a great simplification.~\cite{Chakravarty:2008} However current-current interactions, if relevant, are entirely different matter~\cite{Holstein:1973,Chakravarty:1995}.

The essential problem is now  sewed up. The remainder  can be treated in the Fermi liquid language, perhaps in the formalism of Luttinger-Ward functional,~\cite{Luttinger:1961} if necessary. In particular, broken symmetry states~\cite{Sebastian:2015,Chakravarty:2008b,Millis:2007} can be developed by starting with the mean field starting point. The essential aspects of impurity scattering, both elastic and inelastic, however difficult, can be understood in the Fermi liquid language. 

I was supported by funds from the David S. Saxon Presidential Term Chair at UCLA.  This work was also partly performed at the Aspen Center for Physics, which is supported by National Science Foundation grant PHY-1607611. I would like to thank Chaitanya Murthy and Michael Mulligan for comments.

%\bibliography{QO-23.bib}
%apsrev4-2.bst 2019-01-14 (MD) hand-edited version of apsrev4-1.bst
%Control: key (0)
%Control: author (8) initials jnrlst
%Control: editor formatted (1) identically to author
%Control: production of article title (0) allowed
%Control: page (0) single
%Control: year (1) truncated
%Control: production of eprint (0) enabled
%

\end{document}